\documentclass[journal]{IEEEtran}
\usepackage{amsmath,amsfonts}
\usepackage{array}
\usepackage[caption=false,font=normalsize,labelfont=sf,textfont=sf]{subfig}
\usepackage{textcomp}
\usepackage{stfloats}
\usepackage{url}
\usepackage{verbatim}
\usepackage{graphicx}
\usepackage{cite}
\usepackage{amsmath,amssymb,amsthm}
\usepackage{algorithm,algpseudocode}
\usepackage{booktabs}
\usepackage{graphicx}
\usepackage[colorlinks=true,citecolor=blue]{hyperref}

\newtheorem{definition}{Definition}
\newtheorem{proposition}{Proposition}

\newtheorem{corollary}{Corollary}

\newcommand{\X}{\mathbb{X}}
\newcommand{\LL}{\mathbb{L}}
\newcommand{\R}{\mathbb{R}}
\newcommand{\Z}{\mathbb{Z}}
\newcommand{\cF}{\mathcal{F}}
\newcommand{\cL}{\mathcal{L}}
\newcommand{\cN}{\mathcal{N}}
\newcommand{\bpi}{\boldsymbol{\pi}}
\newcommand{\bfk}{\mathbf{f}}
\newcommand{\Exp}{\mathrm{E}}
\DeclareMathOperator{\cov}{cov}

\begin{document}

\title{Kernel Density Estimation on Labeled Random Finite Sets by Stochastic Imputation for Tracking Swarms}

\title{Multi-Target Tracking of Cooperating Swarms by an Ensemble Gaussian Mixture Filter}
\author{Zain~Jabbar and Andrey~A.~Popov%
\thanks{Z. Jabbar is with the Department of Mathematics, University of Hawai\textquoteleft{}i at M\={a}noa, 2565 McCarthy Mall, Honolulu, HI 96822, USA (e-mail: \mbox{zjabbar@hawaii.edu}; ORCID: 0009-0003-2727-2313).}%
\thanks{A. A. Popov is with the Department of Information and Computer Sciences, University of Hawai\textquoteleft{}i at M\={a}noa, 1680 East-West Road, Honolulu, HI 96822, USA (e-mail: \mbox{apopov@hawaii.edu}; ORCID: 0000-0002-7726-6224).}%
\thanks{Corresponding author: Z. Jabbar. Principal corresponding author: A. A. Popov.}}


\maketitle

\begin{abstract}
  Per-target labeled random-finite-set filters, such as the state of the art $\delta$-generalized labeled multi-Bernoulli filter, cannot represent cooperating, swarm-like, motion between targets.
  In this work, we present a new filter that can.
  The new particle filter couples the joint particle filter with ensemble Gaussian mixture filter such that a measurement updates the density at every particle instead of merely reweighting.
  We prove that the new particle filter converges to the joint particle filter in the limit of particle number and thus to the true Bayesian posterior, under certain assumptions.
  Experimental results on two coupled systems, the Vicsek model and coupled Brownian motion provide evidence on the efficacy of the filter.  
\end{abstract}


\section{Introduction}

State-of-the-art multi-target tracking filters assume that the motion of the targets is independent of one another, which is an assumption that is not necessary in the multi-target Bayes recursion~\cite{mahler2007}.
This work focuses on generalizing this assumption while retaining the standard multi-object measurement model of~\cite{challa2011, vo2024overview}.
Under this measurement model, measurements arrive without identity, targets can be missed, and clutter can be present, so the measurement-to-target association is generally unknown whenever a target is detected~\cite{barshalom2001, mahler2007}.
Targets swarms are a regime in which the independent motion assumption does not hold: the targets cooperate with each other, and commonly occurs in drone fleets~\cite{vo2024overview}, animal collectives~\cite{khan_balch_dellaert_2005}, and satellite constellations~\cite{thomas2018}. 
Because each agent's motion can depend on its neighbors' states~\cite{vicsek1995, cucker_smale_2007}, the multi-target posterior does not in general factor across targets.
A single measurement therefore contains information about every target that is dependent on the associated track, and not just about that one target alone.

We take as the state-of-the-art approach to multi-target tracking the $\delta$-generalized labeled multi-Bernoulli ($\delta$-GLMB) filter~\cite{vo_vo_2013, vo2014labeled, vo2017}, a solution of the multi-target Bayes recursion based on the family of $\delta$-GLMB densities.
Under the standard multi-object measurement model and independent per-target motion~\cite[eq.~(15.23)]{mahler2014}, both the prediction and update steps return a $\delta$-GLMB density~\cite{vo_vo_2013}.
The same family handles the standard difficulties of tracking: clutter, missed detections, unknown and time-varying cardinality, and track management. 
Mahler calls the resulting filter the first tractable, provably Bayes-optimal multi-target detection and tracking algorithm for the standard multi-target motion and measurement models~\cite[pp.~436,~444]{mahler2014}.
The authors of this work contend that this is only true under the assumption of independent per-target motion, and that an alternative to $\delta$-GLMB for this scenario is required.

In this work we assume a particle representation of a probability density, which imposes no functional form on the density it represents. 
Each particle is one realization of the state, and the empirical measure over particles approximates the priors and posteriors.
In the single-target case the particles are vectors in one Euclidean space, and therefore a particle filter like bootstrap particle filter~\cite[p.~126]{sarkka2013} can be readily generalized to more sophisticated density representations by use of the Gaussian mixture models to the ensemble Gaussian mixture filter~\cite{anderson1999monte,yun2022,popov2022}.
In the multi-target case different particles need not share a common cardinality or label set~\cite{khan_balch_dellaert_2005}, so the standard generalization to Gaussian mixtures through the empirical mean and covariance do not apply directly, and thus must be modified in order to be compatible with this formulation.

On a Euclidean state space a point-mass ensemble is expensive in high dimension: to avoid weight collapse the sample size must grow super-exponentially in the effective dimension~\cite[p.~319]{bickel2008}, \cite{bengtsson2008}. 
A joint ensemble of cardinality $L$ can be represented on exactly such a space of dimension $d\cdot L$.
Khan et al.~\cite{khan_balch_dellaert_2005} report the practical form of the cost: a straightforward joint (bootstrap) particle filter is unusable beyond three or four targets, though in the authors' experience this estimate does not take into account advances in computational power.
It has been observed that, while it does not remove the scaling, kernel density estimation at the heart of the ensemble Gaussian mixture filter and the ensemble Epanechnikov mixture filter, for the single-target case, does significantly reduce the number of particles required for the same error~\cite{durant2023,popov2024}.
In labeled random-finite-set (LRFS) multi-target tracking the EnGMF has so far been applied only within a GLMB's single-track factors~\cite{reifler2021}, and not to the entire multi-target state. 

The current work addresses this research gap by introducing the the stochastically imputed labeled random finite set ensemble Gaussian mixture filter (SI-LRFS-EnGMF) of which combines the joint particle filter of Khan et al.~\cite{khan_balch_dellaert_2005}, written for labeled random finite sets, with the kernel density estimate of the ensemble Gaussian-mixture filter (EnGMF).
Stochastic imputation is used to project each particle to a point that lives in $\R^{d\cdot L}$ by drawing the states of its absent labels, so one kernel density estimate covers particles with different label sets.
The cross-target blocks of the kernel covariance are the sample cross-covariances of the states of different targets, and the Kalman gain updates the state of an undetected target through those blocks from the measurements of the targets correlated with it.
This work compares the proposed filter with the joint particle filter and with the $\delta$-GLMB-EnGMF (the $\delta$-GLMB filter with the EnGMF as its per-track filter~\cite{reifler2021}) and the joint particle filter on the Vicsek model~\cite{vicsek1995} and on a swarm Brownian motion.

\section{Background}\label{sec:background}
This section provides background on labeled random finite sets (LRFS), Bayes recursion, the GLMB family, particle filters, and the ensemble Gaussian mixture filter.

\subsection{The LRFS state space}\label{sec:state_space}
In the LRFS state space, each single-object state $x$ in the state space $\X = \R^d$ is augmented with a label $\ell$ from a discrete label space $\LL$~\cite[Sec.~II-A]{vo2014labeled}.
For the purposes of this work $\LL$ is a finite set of $L = |\LL|$ labels.

The definitions of this paragraph are those of~\cite[Secs.~II, II-A, Def.~1]{vo2014labeled}.
The class of finite subsets of a space $\mathbb{S}$ is $\cF(\mathbb{S})$.
The inclusion function is $1_Y(X) = 1$ if $X \subseteq Y$ and $0$ otherwise, with $1_Y(x) = 1_Y(\{x\})$.
The generalized Kronecker delta is $\delta_Y(X) = 1$ if $X = Y$ and $0$ otherwise.
A labeled multi-object state is a finite set $\mathbf{X} \in \cF(\X \times \LL)$ with distinct labels.
Its label set is $\cL(\mathbf{X}) = \{\ell : (x, \ell) \in \mathbf{X}\}$, and the distinct-label indicator $\Delta(\mathbf{X})$ equals one exactly when $|\cL(\mathbf{X})| = |\mathbf{X}|$.
A labeled RFS is a random finite set whose realizations are labeled multi-object states.
For a real-valued function $\chi$ on $\X \times \LL$ the multi-object exponential is $[\chi]^{\mathbf{X}} = \prod_{\mathbf{x} \in \mathbf{X}} \chi(\mathbf{x})$, with $[\chi]^{\emptyset} = 1$.

The set integral of a function $\bfk \colon \cF(\X \times \LL) \to \R$ is
\begin{equation}\label{eq:set_integral}
\int \bfk(\mathbf{X})\, \delta\mathbf{X} \;=\; \sum_{i=0}^{\infty} \frac{1}{i!} \int \bfk(\{\mathbf{x}_1, \ldots, \mathbf{x}_i\})\, d(\mathbf{x}_1, \ldots, \mathbf{x}_i),
\end{equation}
where each $\mathbf{x} = (x, \ell)$ is integrated over $\X \times \LL$ by integrating $x$ over $\X$ and summing $\ell$ over $\LL$~\cite[Sec.~II-C]{vo2014labeled}.
A multi-object density is a nonnegative $\bpi$ whose set integral is one~\cite[Sec.~II-C]{vo2014labeled}.

\subsection{The Bayes recursion and the standard models}\label{sec:bayes_recursion}

Let $\bpi^t_k$ be the multi-object filtering density at time $k$ and $\bpi^t_{k|k-1}$ the prediction density to time $k$; both are conditioned on the measurement history, which the notation leaves implicit~\cite[Sec.~II-B]{vo2014labeled}.
The superscript $t$ marks the true densities of the Bayes recursion~\cite[Sec.~2]{popov2022}, and the superscripts $-$ and $+$ mark a filter's prior and posterior approximations to the true densities.
The multi-object Bayes filter propagates the filtering density by
\begin{align}
\bpi^t_{k+1|k}(\mathbf{X}_{k+1}) &= \int \bfk_{k+1|k}(\mathbf{X}_{k+1}|\mathbf{X}_k)\bpi^t_k(\mathbf{X}_k)\delta\mathbf{X}_k, \label{eq:bayes_predict} \\[2pt]
\bpi^t_{k}(\mathbf{X}_k) &= \frac{g_{k}(Z_{k}|\mathbf{X}_k)\bpi^t_{k|k-1}(\mathbf{X}_k)}{\int g_{k}(Z_{k}|\mathbf{X})\bpi^t_{k|k-1}(\mathbf{X})\delta\mathbf{X}}, \label{eq:bayes_update}
\end{align}
where $\bfk_{k+1|k}$ is the multi-object transition density, $g_k$ the multi-object likelihood, and the integrals are set integrals~\eqref{eq:set_integral}~\cite[eqs.~(1)--(2)]{vo2014labeled}.
The time index is omitted where it is the same throughout a statement, as in~\cite[Sec.~II-B]{vo2014labeled}.

In the standard transition model~\cite[eqs.~(5)--(8)]{vo2014labeled}, each element $(x, \ell)$ of the state $\mathbf{X}_k$ survives to time $k+1$ with probability $p_S(x, \ell)$ and otherwise dies.
A survivor keeps its label and moves to $(x_{k+1}, \ell)$ with density $f(x_{k+1} \mid x, \ell)$.
New objects form a labeled set $\mathbf{Y}$ on a disjoint label space $\mathbb{B}$ with the birth density~\cite[eq.~(5)]{vo2014labeled}
\begin{equation}\label{eq:birth_density}
\bfk_B(\mathbf{Y}) \;=\; \Delta(\mathbf{Y})\, w_B\big(\cL(\mathbf{Y})\big)\, [p_B]^{\mathbf{Y}},
\end{equation}
where $w_B$ is a probability law on $\cF(\mathbb{B})$ and $p_B(\cdot, \ell)$ is the state density of a newborn object with label $\ell$.
The surviving part of the next state $\mathbf{X}_{k+1}$ is $\mathbf{W} = \mathbf{X}_{k+1} \cap (\X \times \cL(\mathbf{X}_k))$.
Vo, Vo, and Phung assume that the objects move independently of one another and that births are independent of survivors~\cite[Sec.~II-B]{vo2014labeled}.
Under those assumptions the multi-object transition kernel is~\cite[eqs.~(6)--(8)]{vo2014labeled}
\begin{align}
\bfk(\mathbf{X}_{k+1} \mid \mathbf{X}_k) &= \bfk_S(\mathbf{W} \mid \mathbf{X}_k)\, \bfk_B(\mathbf{X}_{k+1} - \mathbf{W}), \label{eq:standard_transition}\\
\bfk_S(\mathbf{W} \mid \mathbf{X}) &= \Delta(\mathbf{W})\, \Delta(\mathbf{X})\, 1_{\cL(\mathbf{X})}\big(\cL(\mathbf{W})\big)\, [\Phi(\mathbf{W}; \cdot)]^{\mathbf{X}}, \notag\\
\Phi(\mathbf{W}; x, \ell) &=
\begin{cases}
p_S(x, \ell)\, f(x_{k+1} \mid x, \ell), & (x_{k+1}, \ell) \in \mathbf{W},\\[2pt]
1 - p_S(x, \ell), & \ell \notin \cL(\mathbf{W}).
\end{cases} \notag
\end{align}

For measurements, the sensor reports a finite set $Z = \{z_1, \ldots, z_{|Z|}\} \in \cF(\Z)$ of points of the observation space $\Z = \R^{d_z}$ with no identity information~\cite[Sec.~12.3.5]{mahler2007}.
In the standard measurement model~\cite[p.~419, eq.~(12.121)]{mahler2007}, each element $(x, \ell) \in \mathbf{X}$ is detected with probability $p_D(x, \ell)$, independently of the other elements, and a detected element produces one point of $Z$ with density $g(\cdot \mid x, \ell)$.

For $I \subseteq \LL$ the association maps on $I$ are the functions $\theta \colon I \to \{0, 1, \ldots, |Z|\}$ with $\theta(\ell) = \theta(\ell') > 0$ only if $\ell = \ell'$, and $\Theta(I)$ is the set of them~\cite[Def.~2]{vo2014labeled}.
Here $\theta(\ell) = j$ means $z_j$ is $\ell$'s detection and $\theta(\ell) = 0$ that $\ell$ went undetected.
The likelihood of $Z$ is a sum over these maps, and each term is a product over $\mathbf{X}$ of the factors~\cite[eq.~(4)]{vo2014labeled}
\begin{equation}\label{eq:psi}
\psi_Z(x, \ell; \theta) \;=\;
\begin{cases}
p_D(x, \ell)\, g\big(z_{\theta(\ell)} \mid x, \ell\big), & \theta(\ell) > 0,\\[2pt]
1 - p_D(x, \ell), & \theta(\ell) = 0,
\end{cases}
\end{equation}
without the clutter intensity $\kappa$ of~\cite[eq.~(4)]{vo2014labeled}, since the sensors of Section~\ref{sec:experiments} produce no false alarms~\cite[Sec.~12.3.6]{mahler2007}.
Every reported point is some element's detection~\cite[Sec.~12.3.5]{mahler2007}, so for $I \subseteq \LL$ only the maps in $\Theta_Z(I) = \{\theta \in \Theta(I) : |\{\ell \in I : \theta(\ell) > 0\}| = |Z|\}$ contribute, and the likelihood of the model with no false alarms~\cite[pp.~418--419, eqs.~(12.119)--(12.120)]{mahler2007} is
\begin{equation}\label{eq:mt_likelihood}
g(Z \mid \mathbf{X}) \;=\; \sum_{\theta \in \Theta_Z(\cL(\mathbf{X}))} [\psi_Z(\cdot; \theta)]^{\mathbf{X}}.
\end{equation}
When $|Z| > |\cL(\mathbf{X})|$ no map in $\Theta(\cL(\mathbf{X}))$ reaches every measurement index, so $\Theta_Z(\cL(\mathbf{X}))$ is empty and $g(Z \mid \mathbf{X}) = 0$.

With Poisson false alarms of intensity $\kappa$~\cite[Sec.~II-C]{vo2014labeled}, the likelihood is a factor independent of $\mathbf{X}$ times a sum over $\Theta(\cL(\mathbf{X}))$ of products $[\psi_Z(\cdot; \theta)]^{\mathbf{X}}$ whose detection case~\eqref{eq:psi} is divided by $\kappa(z_{\theta(\ell)})$~\cite[eqs.~(3)--(4)]{vo2014labeled}.

\subsection{The GLMB family}\label{sec:glmb_family}

A generalized labeled multi-Bernoulli (GLMB) density is a mixture of multi-object exponentials indexed by a discrete set~\cite[Sec.~II-D]{vo2014labeled}.
The $\delta$-GLMB form indexes the mixture by pairs of a label set $I$ and a discrete index $\xi$,
\begin{equation}\label{eq:glmb}
\bpi(\mathbf{X}) = \Delta(\mathbf{X}) \sum_{(I, \xi) \in \cF(\LL) \times \Xi} \omega^{(I, \xi)} \, \delta_I(\cL(\mathbf{X}))\, [p^{(\xi)}]^{\mathbf{X}},
\end{equation}
where $\Xi$ is a discrete space and each $p^{(\xi)}(\cdot, \ell)$ is a single-object density on $\X$~\cite[eq.~(9)]{vo2014labeled}.
The weights $\omega^{(I, \xi)} \geq 0$ sum to one over $\cF(\LL) \times \Xi$.
In filtering, $\Xi$ is the space of association-map histories $\xi = (\theta_0, \ldots, \theta_k)$, the pair $(I, \xi)$ is a hypothesis, and $\omega^{(I, \xi)}$ is its probability~\cite[Sec.~II-E]{vo2014labeled}.

Vo and Vo~\cite{vo_vo_2013}, \cite[Props.~3--4]{vo2014labeled} prove that the update~\eqref{eq:bayes_update} under the likelihood~\eqref{eq:mt_likelihood} and the prediction~\eqref{eq:bayes_predict} under the kernel~\eqref{eq:standard_transition} each map a GLMB density to a GLMB density.
The proof of the prediction closure applies each label's own transition $f(\cdot \mid x, \ell)$ to that label's density $p^{(\xi)}(\cdot, \ell)$~\cite[eq.~(22)]{vo2014labeled}.
A kernel whose kinematic factor is one joint density over the surviving labels has no per-label transition, so that proof does not apply to such a kernel.

Vo, Vo, and Phung represent each $p^{(\xi)}(\cdot, \ell)$ by a Gaussian mixture~\cite[Secs.~IV-B, V-B]{vo2014labeled}, and Reifler et al.~\cite{reifler2021} by the EnGMF of Yun, Zanetti, and Jones~\cite{yun2022}.

\subsection{Particle filters}\label{sec:lrfs_kde}

The bootstrap particle filter~\cite[p.~126]{sarkka2013} approximates the filtering density $\pi^t_k$ of~\eqref{eq:bayes_predict}--\eqref{eq:bayes_update} for a single-object state $x \in \R^d$ by the empirical measure of $N$ samples $x^{(1)}, \ldots, x^{(N)}$.
One time step is
\begin{align}
x^{(i)-} &\sim f(\cdot \mid x^{(i)}), \label{eq:bootstrap_step}\\
w^{(i)} &\propto g(z \mid x^{(i)-}), \label{eq:bootstrap_weight}\\
a(i) &\overset{\text{i.i.d.}}{\sim} \mathrm{Categorical}\big(w^{(1)}, \ldots, w^{(N)}\big), \label{eq:bootstrap_ancestor}\\
x^{(i)} &= x^{(a(i))-}, \label{eq:bootstrap_resample}
\end{align}
with $x^{(i)-}$ the predicted sample and $a \colon \{1, \ldots, N\} \to \{1, \ldots, N\}$ the map of ancestor variables of the resampling~\cite[Algorithm~2]{gerber2019}.

The multi-object density $\bpi$ of Section~\ref{sec:state_space} is a density with respect to the set integral~\eqref{eq:set_integral}, so~\eqref{eq:bootstrap_step}--\eqref{eq:bootstrap_resample} apply to $\bpi$ mutatis mutandis~\cite{vo_singh_doucet_2005}.
The changes are $\mathbf{X}$ in place of $x$, the multi-object transition $\bfk$ of~\eqref{eq:bayes_predict} in place of the single-object one, and the likelihood~\eqref{eq:mt_likelihood} in place of $g$.
Because $\bfk$ is sampled and never evaluated~\cite[Ch.~7]{sarkka2013}, its kinematic factor may be one joint density over the surviving labels.

Orton and Fitzgerald~\cite[eq.~(7), Algorithm~2]{orton2002bayesian} fix the number of targets and propose from a Gaussian approximation to the optimal importance function~\cite[Ch.~7]{sarkka2013} instead of from $\bfk$.
Khan, Balch, and Dellaert~\cite[eqs.~(6), (12)]{khan_balch_dellaert_2005} fix the label set and put the interaction between targets in $\bfk$ as pairwise Markov-random-field potentials.
Yang et al.~\cite{yang2024} track clusters of coupled agents with a continuous-discrete Poisson multi-Bernoulli mixture filter~\cite{garcia_fernandez_2018} on a joint stochastic differential equation.
The filter of Section~\ref{sec:swarm_lrfs_filter} samples a transition kernel with dependence between targets, as~\cite{khan_balch_dellaert_2005} and~\cite{yang2024} do, and its update also uses the cross-target covariance of the forecast ensemble.

\subsection{The EnGMF}\label{sec:engmf_background}

A kernel density estimate on $\R^d$ places one kernel at each sample~\cite[Sec.~4.2]{silverman1986}.
The ensemble Gaussian-mixture filter (EnGMF) places at each forecast sample a Gaussian kernel whose covariance, the bandwidth, is shared across samples~\cite[eqs.~(2.6)--(2.7)]{popov2022}.

A state $x \in \R^d$ is observed as~\cite[eq.~(2.2)]{popov2022}
\begin{equation}\label{eq:nonlinear_measurement}
z \;=\; h(x) + \eta, \qquad \eta \sim \cN(0, R),
\end{equation}
through a differentiable map $h$, with the Gaussian likelihood $g(z \mid x) = \cN(z; h(x), R)$.
The forecast samples $x^{(1)}, \ldots, x^{(N)} \in \R^d$ are the previous posterior samples propagated through the dynamics~\cite[eqs.~(2.1), (2.7)]{popov2022}.
Let $B$ be a bandwidth.
The linearization of $h$ about sample $i$ is $H^{(i)} = \nabla h(x^{(i)})$~\cite[p.~3]{popov2022}.
The forecast density is the mixture $\frac{1}{N} \sum_{i} \cN(x; x^{(i)}, B)$.
The update linearizes $h$ at each sample~\cite[eq.~(2.9)]{popov2022}, which gives the posterior mixture
\begin{align}
p^+(x|z) &= \sum_{i=1}^{N} w^{(i)} \cN(x; m^{(i)}, P^{(i)}), \label{eq:engmf_update}\\
m^{(i)} &= x^{(i)} + K^{(i)}(z - h(x^{(i)})), \label{eq:engmf_mean}\\
P^{(i)} &= (I_d - K^{(i)} H^{(i)}) B, \label{eq:engmf_cov}\\
K^{(i)} &= B H^{(i)T} (S^{(i)})^{-1}, \label{eq:engmf_gain}\\
S^{(i)} &= H^{(i)} B H^{(i)T} + R, \label{eq:engmf_innovation}\\
w^{(i)} &\propto \cN(z; h(x^{(i)}), S^{(i)}), \label{eq:engmf_weight}
\end{align}
with $m^{(i)}$, $P^{(i)}$, $K^{(i)}$, and $S^{(i)}$ the posterior mean, posterior covariance, gain, and innovation covariance of the Kalman update~\cite[Sec.~5.2.3]{barshalom2001}, and $w^{(i)}$ the component's evidence~\cite[eq.~(2.10)]{popov2022}.
When $h(x) = Hx$ the matrices $H^{(i)}$, $P^{(i)}$, $K^{(i)}$, and $S^{(i)}$ are the same for every sample and~\eqref{eq:engmf_update} is the exact posterior of the mixture~\cite[p.~3]{popov2022}.
The next ensemble is $N$ i.i.d.\ draws from~\eqref{eq:engmf_update}.

\section{The SI-LRFS-EnGMF}\label{sec:swarm_lrfs_filter}

The defining feature of the proposed algorithm is that it projects the state space of the LRFS into Euclidean space with a maximum of $L$ possible targets.

Algorithm~\ref{alg:swarm_lrfs} is one iteration of the recursion~\eqref{eq:bayes_predict}--\eqref{eq:bayes_update} on the particles of Section~\ref{sec:lrfs_kde}, with the kernel step of Section~\ref{sec:engmf_background} in place of the reweighting~\eqref{eq:bootstrap_weight}, which is now described.
A particle $\mathbf{X}^{(i)}$ has label set $I^{(i)} = \cL(\mathbf{X}^{(i)})$, and $\Lambda^{(i)} = \Lambda(\mathbf{X}^{(i)}) \in \R^{dL}$ is the image of $\mathbf{X}^{(i)}$ under the imputation map of Definition~\ref{def:stochastic_imputation}.
The filter's prior belief $\bpi^-_k$ and posterior belief $\bpi^+_k$ are the instances of Definition~\ref{def:belief} that approximate $\bpi^t_{k|k-1}$ and $\bpi^t_k$.

\begin{algorithm}[t]
\caption{SI-LRFS-EnGMF, one iteration.}\label{alg:swarm_lrfs}
\begin{algorithmic}[1]
\Require particles $\{\mathbf{X}^{(i)}\}_{i=1}^{N}$; $Z$; the joint kinematic transition of $\bfk$; $(H, R)$; $p_D$; $n_0$; $(p_S, r_B, n_B, m_B, P_B)$
\Statex \textbf{Predict} \Comment{Section~\ref{sec:filter}}
\State $\mathbf{W}^{(i)} :=$ each $(x, \ell) \in \mathbf{X}^{(i)}$ kept with probability $p_S$, its state $x$ moved by the joint transition of $\bfk$
\State $\mathbf{X}^{(i)} := \mathbf{W}^{(i)} \cup \mathbf{Y}^{(i)}$;\quad $\mathbf{Y}^{(i)} \sim \bfk_B$ of~\eqref{eq:birth_density} with $w_B$ given by~\eqref{eq:birth_death_kernel} and $p_B(\cdot, \ell) = \cN(\cdot;\; m_B, P_B)$
\Statex \textbf{Update}
\State $\mathbf{X}^{(i)} := \sigma_i \cdot \mathbf{X}^{(i)}$;\quad $I^{(i)} := \cL(\mathbf{X}^{(i)})$ \Comment{Section~\ref{sec:slot_alignment}}
\State $\Lambda^{(i)} := \Lambda(\mathbf{X}^{(i)})$ \Comment{Definition~\ref{def:stochastic_imputation}}
\State $B := \beta_N\, \widehat{\cov}\big(\{\Lambda^{(i)}\}_{i=1}^{N}\big)$ \Comment{\eqref{eq:silverman}}
\For{$i = 1, \ldots, N$ in parallel}
    \State $\theta^{(i)\star} := \arg\max_{\theta \in \Theta^\star_Z(I^{(i)})} \sum_{\ell \in I^{(i)}} \log \psi_Z\big(\Lambda^{(i)}_\ell, \ell; \theta\big)$ \Comment{\eqref{eq:per_particle_cost}}
    \State $w^{(i, \theta^{(i)\star})} :=$~\eqref{eq:point_weight}
    \State $m^{(i, \theta^{(i)\star})} := \Lambda^{(i)} + K^{(i, \theta^{(i)\star})} s^{(i, \theta^{(i)\star})}$ \Comment{\eqref{eq:mixture_posterior}}
\EndFor
\Statex \textbf{Resample} \Comment{\eqref{eq:resample}}
\State $a(i) := \min\{j : F(j) \ge (i - 1 + u_i)/N\}$, $u_i \overset{\text{i.i.d.}}{\sim} \mathrm{Unif}[0, 1)$
\State $v^{(i)} \sim \cN\big(\cdot;\; m^{(a(i), \theta^{(a(i))\star})},\; P^{(a(i), \theta^{(a(i))\star})}\big)$;\quad $\mathbf{X}^{(i)} := \{(v^{(i)}_\ell, \ell) : \ell \in I^{(a(i))}\}$
\State \Return $\{\mathbf{X}^{(i)}\}_{i=1}^{N}$
\end{algorithmic}
\end{algorithm}

The labels $\ell_1, \ldots, \ell_L$ of $\LL$ are taken in a fixed order, which identifies the product space $\X^\LL$ with $\R^{dL}$,
\begin{equation}\label{eq:joint_state}
\X^\LL \;=\; \X \times \cdots \times \X \;\cong\; \R^{dL},
\end{equation}
and a vector $v \in \R^{dL}$ has the block $v_\ell \in \X$ at label $\ell$ and the blocks $v_I \in \X^I$ at a label set $I \subseteq \LL$; a matrix of $\R^{dL \times dL}$ has blocks $P_{\ell \ell'}$ and $P_{II}$ in the same way.

The kernel density estimate of Section~\ref{sec:engmf_background} needs one kernel center per particle, all in one Euclidean space, because its bandwidth is a scalar multiple of the sample covariance of the centers.
Under births and deaths a particle with label set $I \subsetneq \LL$ determines the blocks at $I$ only and no vector of $\R^{dL}$.
Label sets of different cardinality give vectors of different length, between which no sample covariance is defined.
Distinct label sets of equal cardinality give vectors of equal length whose coordinates belong to different labels, so a sample covariance computed entrywise compares states that no common label indexes.

\subsection{Stochastic imputation}\label{sec:stochastic_imputation}

\begin{definition}[Stochastic imputation]\label{def:stochastic_imputation}
Fix one probability density $\tilde p_\ell$ on $\X$ for each $\ell \in \LL$; together they constitute the imputation law.
Stochastic imputation is the random map $\Lambda$ from $\{\mathbf{X} \in \cF(\X \times \LL) : \Delta(\mathbf{X}) = 1\}$ to $\R^{dL}$ whose value at $\mathbf{X}$ has, at each $\ell \in \LL$, the block
\begin{equation*}
\Lambda(\mathbf{X})_\ell \;=\;
\begin{cases}
x, & (x, \ell) \in \mathbf{X}, \\[2pt]
\varepsilon_\ell \sim \tilde p_\ell \ \text{independently}, & \ell \notin \cL(\mathbf{X}).
\end{cases}
\end{equation*}
For $I \subseteq \LL$, $\Lambda(\mathbf{X})_I \in \R^{d|I|}$ is the vector of the blocks $\Lambda(\mathbf{X})_\ell$ for $\ell \in I$, which does not depend on the draws $\varepsilon$ when $I \subseteq \cL(\mathbf{X})$.
\end{definition}

Let $p_B(\cdot, \ell) = \cN(\cdot;\; m_B, P_B)$ be the newborn state density of~\eqref{eq:birth_density} for every $\ell$, $n_0 > 0$ a fixed constant, and $n_\ell = |\{i : \ell \in I^{(i)}\}|$.
For $n_\ell > 0$,
\begin{align}
m_\ell &= \frac{1}{n_\ell} \sum_{i : \ell \in I^{(i)}} \Lambda^{(i)}_\ell, \label{eq:label_mean}\\
P_\ell &= \frac{1}{n_\ell} \sum_{i : \ell \in I^{(i)}} (\Lambda^{(i)}_\ell - m_\ell)(\Lambda^{(i)}_\ell - m_\ell)^T, \label{eq:label_cov}
\end{align}
and $n_\ell m_\ell = 0$ and $n_\ell P_\ell = 0$ when $n_\ell = 0$.
Each $\tilde p_\ell$ is the Gaussian $\cN(\cdot;\; \tilde m_\ell, \tilde P_\ell)$ with
\begin{align}
\tilde m_\ell &= \frac{n_\ell\, m_\ell + n_0\, m_B}{n_\ell + n_0}, \label{eq:blend_imputation}\\
\tilde P_\ell &= \frac{n_\ell\, P_\ell + n_0\, P_B}{n_\ell + n_0}. \label{eq:blend_imputation_cov}
\end{align}

\subsection{Label alignment}\label{sec:slot_alignment}

For a permutation $\sigma$ of $\LL$ the relabeled state is $\sigma \cdot \mathbf{X} = \{(x, \sigma(\ell)) : (x, \ell) \in \mathbf{X}\}$.
The posterior of every model in this paper is invariant under relabeling, $\bpi^t_k(\sigma \cdot \mathbf{X}) = \bpi^t_k(\mathbf{X})$ for every $\sigma$, the invariance that causes the label-switching problem of Bayesian mixture posteriors~\cite{stephens2000}.
The invariance holds at every $k$ by induction on~\eqref{eq:bayes_predict}--\eqref{eq:bayes_update}, because the initial density, $p_S$, the joint transition, the birth law~\eqref{eq:birth_death_kernel}, $p_D$, and $g(\cdot \mid x, \ell)$ do not depend on the label and $\mathbf{X} \mapsto \sigma \cdot \mathbf{X}$ preserves the set integral~\eqref{eq:set_integral}.

Alignment is the relabeling algorithm of Stephens~\cite{stephens2000} run once per time step on the labeled representation.
Each particle is replaced by a relabeled copy of itself, $\mathbf{X}^{(i)} \mapsto \sigma_i \cdot \mathbf{X}^{(i)}$, with $\sigma_i$ computed from the ensemble alone.
Relabeling changes neither the density of a particle nor the law of a label-invariant function of one realization, such as its cardinality or its optimal sub-pattern assignment (OSPA) distance to the truth~\cite{schuhmacher2008}.

Each label's reference is the imputation mean $\tilde m_\ell$ of~\eqref{eq:blend_imputation}.
The permutation $\sigma_i$ is a minimizer of $\sum_{\ell \in \LL} c^{(i)}(\ell, \sigma(\ell))$ over permutations, a minimum-cost assignment~\cite{kuhn1955} with cost
\begin{equation}\label{eq:alignment_cost}
c^{(i)}(\ell, \ell') \;=\;
\begin{cases}
\big\| \Lambda^{(i)}_\ell - \tilde m_{\ell'} \big\|^2, & \ell \in I^{(i)}, \\[2pt]
n_{\ell'} / N, & \ell \notin I^{(i)}.
\end{cases}
\end{equation}
References and assignments are recomputed alternately for a fixed number of passes, which is Algorithm~1 of Stephens~\cite{stephens2000} in the squared-error instance he identifies as the batch relabeling of Celeux~\cite{celeux1998}.
The second case of~\eqref{eq:alignment_cost} has no counterpart in a mixture with a fixed number of components, where every label is present in every draw.
Its total over a permutation, $\sum_{\ell' \notin \sigma(I^{(i)})} n_{\ell'} / N$, is least when the absent labels of particle $i$ are placed at the labels of smallest $n_{\ell'}$.

\subsection{The belief}\label{sec:belief}

\begin{definition}[Belief]\label{def:belief}
A belief is a collection of $N$ components $(w^{(i)}, I^{(i)}, m^{(i)}, P^{(i)})$: weights $w^{(i)} \ge 0$ with $\sum_i w^{(i)} = 1$, label sets $I^{(i)} \subseteq \LL$, means $m^{(i)} \in \R^{dL}$, and symmetric positive semidefinite covariances $P^{(i)} \in \R^{dL \times dL}$.
Its density with respect to the set integral~\eqref{eq:set_integral} at $\mathbf{X} \in \cF(\X \times \LL)$ with label set $I = \cL(\mathbf{X})$ is
\begin{equation}\label{eq:belief_mixture}
\bpi(\mathbf{X}) = \Delta(\mathbf{X}) \sum_{i=1}^{N} w^{(i)} \delta_{I^{(i)}}(I) \cN(\Lambda(\mathbf{X})_I; m^{(i)}_I, P^{(i)}_{II}),
\end{equation}
where $m^{(i)}_I$ and $P^{(i)}_{II}$ are the blocks of $m^{(i)}$ and $P^{(i)}$ at $I$ in the ordering of~\eqref{eq:joint_state}.
When $P^{(i)}_{II}$ is singular, $\cN(\cdot;\; m^{(i)}_I, P^{(i)}_{II})$ is the Gaussian measure with that mean and covariance, and~\eqref{eq:belief_mixture} defines a measure rather than a density.
\end{definition}

The existence probability of label $\ell$ under~\eqref{eq:belief_mixture} is $\hat r(\ell) = \sum_{i=1}^{N} w^{(i)}\, 1_{I^{(i)}}(\ell)$, as for the $\delta$-GLMB~\cite[Sec.~VI]{vo2014labeled}.

The entries of $m^{(i)}$ and $P^{(i)}$ outside the blocks of $I^{(i)}$ do not appear in~\eqref{eq:belief_mixture}.
Both are kept on all of $\R^{dL}$ because the update~\eqref{eq:engmf_update} of Section~\ref{sec:filter} is a map on $\R^{dL}$.

One iteration of Algorithm~\ref{alg:swarm_lrfs} passes through three instances of Definition~\ref{def:belief}.
The prior belief $\bpi^-$ has $w^{(i)} = 1/N$, $m^{(i)}_\ell = \Lambda^{(i)}_\ell$ for $\ell \in I^{(i)}$, and $P^{(i)} = 0$, so its Gaussian factor is the point mass at $\Lambda^{(i)}_{I^{(i)}}$ and $\bpi^-$ is the empirical measure~\cite[p.~126]{sarkka2013} of the particles $\{\mathbf{X}^{(i)}\}_{i=1}^{N}$ of Section~\ref{sec:lrfs_kde}.
This instance is the input of Algorithm~\ref{alg:swarm_lrfs}, and the output is the same instance at the next time step.

The kernel estimate is
\begin{equation}\label{eq:gmm_lrfs}
p^-(v) \;=\; \frac{1}{N} \sum_{i=1}^{N} \cN\!\left(v;\; \Lambda^{(i)},\; B\right), \qquad v \in \R^{dL}.
\end{equation}
The components $(1/N, I^{(i)}, \Lambda^{(i)}, B)$ of~\eqref{eq:gmm_lrfs} form an instance of Definition~\ref{def:belief}.
The bandwidth $B$ is Silverman's rule~\cite[Sec.~4.3.2, eq.~(4.14)]{silverman1986} at the joint dimension, the factor
\begin{equation}\label{eq:silverman}
\beta_N = \Big(\tfrac{4}{N(dL + 2)}\Big)^{2/(dL + 4)}
\end{equation}
times the sample covariance of the imputed ensemble, whose $(\ell, \ell')$ block is
\begin{equation}\label{eq:blended_moment}
B_{\ell\ell'} \;=\; \frac{\beta_N}{N - 1} \sum_{i=1}^{N} \big(\Lambda^{(i)}_\ell - \bar\Lambda_\ell\big)\big(\Lambda^{(i)}_{\ell'} - \bar\Lambda_{\ell'}\big)^{T},
\end{equation}
with $\bar\Lambda = \frac{1}{N} \sum_{i=1}^{N} \Lambda^{(i)}$ the mean of the imputed ensemble.
The rank of $B$ is at most $N - 1$, so $B$ is singular when $N \le dL$, and~\eqref{eq:gmm_lrfs} is then a measure in the sense of Definition~\ref{def:belief}.

\subsection{The filter}\label{sec:filter}

Lines 1 and 2 sample the transition~\eqref{eq:standard_transition} with the kinematic factor of $\bfk_S$ replaced by one joint transition density of the surviving states~\cite[Algorithm~1]{khan_balch_dellaert_2005}.
The number $b$ of births and the set $I_B \subseteq \LL - \cL(\mathbf{W})$ of newborn labels have the law $w_B$ of~\eqref{eq:birth_density} given by
\begin{align}
b &\sim \mathrm{Binomial}(n_B, r_B), \label{eq:birth_count}\\
I_B &\sim \mathcal{U}\{J \subseteq \LL - \cL(\mathbf{W}) : |J| = \min(b, L - |\mathbf{W}|)\}. \label{eq:birth_death_kernel}
\end{align}

Lines 7--9 apply the EnGMF update~\eqref{eq:engmf_update} to~\eqref{eq:gmm_lrfs} with, for component $i$, the association map $\theta^{(i)\star}$ of line 7 in the likelihood.
Fix one particle $\mathbf{X} = \mathbf{X}^{(i)}$, with $I = \cL(\mathbf{X})$ and $\Lambda = \Lambda(\mathbf{X})$, and write $\theta^\star$ for $\theta^{(i)\star} \in \Theta(I)$, since the association map and the weight of particle $i$ depend on no other particle.
At $\theta \in \Theta(I)$ the summand of the likelihood~\eqref{eq:mt_likelihood} is $[\psi_Z(\cdot; \theta)]^{\mathbf{X}} = \prod_{\ell \in I} \psi_Z(\Lambda_\ell, \ell; \theta)$.
Let $\Theta^\star_Z(I) = \{\theta \in \Theta(I) : |\{\ell \in I : \theta(\ell) > 0\}| = \min(|Z|, |I|)\}$, a set that is nonempty for every $Z$ and equals $\Theta_Z(I)$ when $|Z| \le |I|$.
Line 7 takes a maximizer $\theta^\star$ over $\Theta^\star_Z(I)$ by the Hungarian algorithm~\cite{kuhn1955}, as the minimum-cost assignment of rows to columns of the $|I| \times (|Z| + |I|)$ matrix $C_Z$ with, for the labels $\ell_1, \ldots, \ell_{|I|}$ of $I$ in the order of~\eqref{eq:joint_state},
\begin{equation}\label{eq:per_particle_cost}
[C_Z]_{\alpha j} =
\left\{\begin{array}{@{}l@{\mkern8mu}l@{}}
-\log[p_D g(z_j|\Lambda_{\ell_\alpha},\ell_\alpha)]-\lambda, & j \le |Z|,\\[2pt]
-\log(1 - p_D), & j = |Z|+\alpha,\\[2pt]
\infty, & \text{otherwise},
\end{array}\right.
\end{equation}
where the penalty $\lambda$ for each measurement left unassigned exceeds the difference of every two values in $\{-\log[p_D g(z_j|\Lambda_{\ell_\alpha}, \ell_\alpha)] : j \le |Z|, \alpha \le |I|\} \cup \{-\log(1 - p_D)\}$.
Row $\alpha$ assigned to column $j \le |Z|$ sets $\theta^\star(\ell_\alpha) = j$, and row $\alpha$ assigned to column $|Z| + \alpha$ sets $\theta^\star(\ell_\alpha) = 0$.
Each column $j \le |Z|$ in an assignment adds $-\lambda$ to its cost, so every minimizer uses $\min(|Z|, |I|)$ such columns and lies in $\Theta^\star_Z(I)$.

Line 8 sets
\begin{equation}\label{eq:point_weight}
w^{(\theta^\star)} \;=\; e^{-(|Z| - \min(|Z|, |I|))\lambda} \prod_{\ell \in I} \psi_Z(\Lambda_\ell, \ell; \theta^\star),
\end{equation}
the summand of the likelihood~\eqref{eq:mt_likelihood} at $\theta^\star$ when $|Z| \le |I|$.
For $|Z| > |I|$ the likelihood~\eqref{eq:mt_likelihood} is zero, and the weight replaces zero by the first factor of~\eqref{eq:point_weight}.

Let $\LL^\theta = \{\ell \in \LL : \theta(\ell) > 0\}$ be the detected labels of $\theta$.
At $\theta$ the Kalman step is~\eqref{eq:engmf_update} on $\R^{dL}$ with the bandwidth $B$ of~\eqref{eq:silverman} and, in place of $(z, h, R)$,
\begin{align}
z^{(\theta)} &= (z_{\theta(\ell)})_{\ell \in \LL^\theta}, \label{eq:joint_measurement}\\
h^{(\theta)}(v) &= (h(v_\ell))_{\ell \in \LL^\theta}, \label{eq:joint_sensor}\\
R^{(\theta)} &= I_{|\LL^\theta|} \otimes R, \label{eq:joint_noise}
\end{align}
with $\otimes$ the Kronecker product.
The Jacobian of $h^{(\theta)}$ at $\Lambda^{(i)}$ has the blocks $H^{(i, \theta)}_{\ell \ell'} = \delta_\ell(\ell')\, \nabla h(\Lambda^{(i)}_\ell)$, with $\ell \in \LL^\theta$ indexing the row blocks in the order of $z^{(\theta)}$ and $\ell' \in \LL$ the column blocks.
The innovation is $s^{(i, \theta)} = z^{(\theta)} - h^{(\theta)}(\Lambda^{(i)})$.
The matrices $S^{(i, \theta)}$, $K^{(i, \theta)}$, and $P^{(i, \theta)}$ are $S$, $K$, and $P$ of~\eqref{eq:engmf_cov}--\eqref{eq:engmf_innovation} with $H^{(i, \theta)}$, $R^{(\theta)}$, and $I_{dL}$ in place of $H^{(i)}$, $R$, and $I_d$.

The mean of component $i$ at $\theta$ is $m^{(i, \theta)} = \Lambda^{(i)} + K^{(i, \theta)} s^{(i, \theta)}$ by~\eqref{eq:engmf_mean}, and its weight is $\bar w^{(i, \theta)} \propto \cN(z^{(\theta)};\; h^{(\theta)}(\Lambda^{(i)}),\; S^{(i, \theta)})$ by~\eqref{eq:engmf_weight}.
The posterior belief at $\theta$ has the density on $\R^{dL}$
\begin{equation}\label{eq:mixture_posterior}
p^+(v \mid Z, \theta)
\;=\; \sum_{i=1}^{N} \bar w^{(i, \theta)} \, \cN\!\left(v;\; m^{(i, \theta)},\; P^{(i, \theta)}\right).
\end{equation}
Line 8 uses the weight $w^{(i, \theta^{(i)\star})}$ of~\eqref{eq:point_weight} in place of $\bar w^{(i, \theta)}$.
The weight $\log w^{(i, \theta)}$ varies with $i$ through a quadratic form in $(R^{(\theta)})^{-1}$ and $\log \bar w^{(i, \theta)}$ through one in $(S^{(i, \theta)})^{-1}$.
By~\eqref{eq:engmf_innovation}, $S^{(i, \theta)} \to R^{(\theta)}$ as $B \to 0$, so the two weights, normalized over $i$, coincide in that limit.

The block row of $K^{(i, \theta)}$ at label $\ell \in \LL$ is
\begin{equation}\label{eq:gain_row}
K^{(i, \theta)}_{\ell} \;=\; \big[\, B_{\ell \ell'}\, \nabla h(\Lambda^{(i)}_{\ell'})^T \,\big]_{\ell' \in \LL^\theta}\; (S^{(i, \theta)})^{-1}.
\end{equation}
Row $\ell$ is zero exactly when $B_{\ell \ell'} \nabla h(\Lambda^{(i)}_{\ell'})^T = 0$ for every $\ell' \in \LL^\theta$, so the posterior mean of an undetected label differs from $\Lambda^{(i)}_\ell$ only through those products.

Lines 11 and 12 draw the next prior belief from the posterior belief $\bpi^+$, the mixture~\eqref{eq:mixture_posterior} with $\theta = \theta^{(i)\star}$ in component $i$, by stratified resampling~\cite[Sec.~2.2]{gerber2019}.
With $F(j) = \sum_{j' \le j} w^{(j', \theta^{(j')\star})} / \sum_{j'=1}^{N} w^{(j', \theta^{(j')\star})}$ the cumulative normalized weights and $u_1, \ldots, u_N$ i.i.d.\ uniform on $[0, 1)$,
\begin{gather}
a(i) = \min\{j : F(j) \ge (i - 1 + u_i)/N\}, \notag\\
v^{(i)} \sim \cN\big(\cdot;\; m^{(a(i), \theta^{(a(i))\star})},\; P^{(a(i), \theta^{(a(i))\star})}\big), \label{eq:resample}\\
\mathbf{X}^{(i)} = \{(v^{(i)}_\ell, \ell) : \ell \in I^{(a(i))}\}. \notag
\end{gather}
Stratified resampling is consistent~\cite[Cor.~2]{gerber2019}.

The update solves one linear system in the $(d_z |\LL^\theta|) \times (d_z |\LL^\theta|)$ matrix $S^{(i, \theta)}$ per particle.
Its cost is $O(N (d_z L)^3)$ floating-point operations in total.

\subsection{Representational capacity}\label{sec:representation}

\begin{proposition}[Containment]\label{prop:containment}
Let $\bpi$ be a $\delta$-GLMB density~\eqref{eq:glmb} with
\begin{equation}\label{eq:gm_single_object}
p^{(\xi)}(x, \ell) \;=\; \sum_{j=1}^{Q_{\xi,\ell}} c_{\xi,\ell,j}\, \cN\big(x;\; m_{\xi,\ell,j},\; P_{\xi,\ell,j}\big)
\end{equation}
for every $(\xi, \ell)$, with $c_{\xi,\ell,j} \ge 0$ and $\sum_j c_{\xi,\ell,j} = 1$.
Then $\bpi$ is the density~\eqref{eq:belief_mixture} with one component per pair $(I, \xi)$ and index tuple $j = (j_\ell)_{\ell \in I}$, of label set $I$, weight $\omega^{(I,\xi)} \prod_{\ell \in I} c_{\xi,\ell,j_\ell}$, mean with block $m_{\xi,\ell,j_\ell}$ at each $\ell \in I$ and zero elsewhere, and covariance with block $P_{\xi,\ell,j_\ell}$ at $(\ell, \ell)$ for $\ell \in I$ and zero elsewhere.
The number of components is $N = \sum_{(I, \xi)} \prod_{\ell \in I} Q_{\xi,\ell}$.
\end{proposition}

\begin{proposition}[Strictness and closure]\label{prop:strictness}
(i) Let $I \subseteq \LL$ with $|I| \ge 2$, and let $p = \cN(\cdot;\; m, P)$ be a Gaussian density on $\R^{d|I|}$ with $P_{\ell\ell'} \ne 0$ for some $\ell \ne \ell'$ in $I$.
Then $p$ is not a finite mixture of Gaussian densities whose covariances are block-diagonal over $I$~\cite[eq.~18.18.28]{dlmf}, \cite[Sec.~5, eq.~(5.1)]{stewart1993}.
(ii) Every density of Definition~\ref{def:belief} is the $L^1$ limit of a sequence of densities of Definition~\ref{def:belief} whose covariances are $\varsigma_j^2 I_{dL}$ with $\varsigma_j > 0$~\cite[Thm.~5(f)]{nguyen2020}.
\end{proposition}
The correlation coefficient of an entry $(P_{\ell\ell'})_{jj'} \ne 0$ is $\varrho = (P_{\ell\ell'})_{jj'} / \sqrt{(P_{\ell\ell})_{jj} (P_{\ell'\ell'})_{j'j'}}$.
The number of Gaussian densities with covariances block-diagonal over $I$ that a mixture needs to come within a fixed $L^2$ distance of $p$ grows without bound as $\varrho \to \pm 1$~\cite[Sec.~5, eqs.~(5.2)--(5.3)]{stewart1993}.

\begin{corollary}[Conditional independence under a $\delta$-GLMB posterior]\label{cor:conditional_independence}
Let the likelihood be a sum over association maps of products $[\psi_Z(\cdot; \theta)]^{\mathbf{X}}$, with $\psi_Z$ as in~\eqref{eq:psi} at any $p_D(x, \ell)$ and any clutter intensity $\kappa$.
Under the $\delta$-GLMB posterior, conditional on a hypothesis $(I, \xi)$, the states $\Lambda(\mathbf{X})_\ell$ and $\Lambda(\mathbf{X})_{\ell'}$ of two labels $\ell \ne \ell'$ in $I$ are independent.
\end{corollary}
The corollary is the update closure~\cite[Sec.~II-F, Prop.~3]{vo2014labeled}, whose updated density on the hypothesis $(I, \xi, \theta)$ is the product $[p^{(\xi, \theta)}]^{\mathbf{X}}$ of single-object densities.
The joint density of the states of two labels under a $\delta$-GLMB posterior is therefore the mixture over $\xi$, with weights $\omega^{(I, \xi)}$, of product densities.

The kernel estimate~\eqref{eq:gmm_lrfs} has the cross-covariance block $B_{\ell\ell'}$ of~\eqref{eq:blended_moment}, the sample cross-covariance of the forecast ensemble, at every $p_D$ and $\kappa$.
By~\eqref{eq:gain_row}, the blocks $B_{\ell\ell'}$ with $\ell' \in \LL^\theta$ move the posterior mean of component $i$ at a label $\ell \notin \LL^\theta$.
In the $\delta$-GLMB-EnGMF each label is updated with the gain of its own block, $B_{\ell\ell} \nabla h(\Lambda^{(i)}_\ell)^T (\nabla h(\Lambda^{(i)}_\ell) B_{\ell\ell} \nabla h(\Lambda^{(i)}_\ell)^T + R)^{-1}$, so within a hypothesis with $\theta(\ell) = 0$ the density of label $\ell$ is its predicted density.
When $\ell \notin I^{(i)}$ the coordinate moved by~\eqref{eq:gain_row} is an imputed one, which the output particle of~\eqref{eq:resample} drops, and the output depends on the imputation draws through $B$ alone, by Proposition~\ref{prop:imputation_inert}(iii).

For a joint density $p$ of the states $\Lambda(\mathbf{X})_\ell$ and $\Lambda(\mathbf{X})_{\ell'}$ of two labels, with marginals $p_\ell$, $p_{\ell'}$, and any product density $q_\ell q_{\ell'}$, the Kullback--Leibler divergence $D_{\mathrm{KL}}(\cdot \,\|\, \cdot)$~\cite[Sec.~2.3]{cover2006} decomposes as
\begin{multline}\label{eq:kl_decomposition}
D_{\mathrm{KL}}(p \,\|\, q_\ell q_{\ell'}) \;=\; D_{\mathrm{KL}}(p \,\|\, p_\ell p_{\ell'}) \\ + D_{\mathrm{KL}}(p_\ell \,\|\, q_\ell) + D_{\mathrm{KL}}(p_{\ell'} \,\|\, q_{\ell'}).
\end{multline}
The product density closest to $p$ is therefore $p_\ell p_{\ell'}$, at divergence equal to the mutual information $D_{\mathrm{KL}}(p \,\|\, p_\ell p_{\ell'})$ of the two states~\cite[eq.~(2.29)]{cover2006}.
For a pair of scalar Gaussian variables with correlation coefficient $\varrho$ that divergence is $-\tfrac{1}{2}\log(1 - \varrho^2)$ nats~\cite[Example~8.5.1]{cover2006}.
Section~\ref{sec:sbm} varies $\varrho$ between the states of two labels through the coupling $\gamma$ of the truth process, and Section~\ref{sec:pd_dependence} varies the number of undetected labels through $p_D$.

\subsection{The large-ensemble limit}\label{sec:convergence}

The idealized filter is Algorithm~\ref{alg:swarm_lrfs} with the weight
\begin{equation}\label{eq:idealised_weight}
\tilde g(Z|\mathbf{X}) = e^{-(|Z| - \min(|Z|, |I|))\lambda} \sum_{\theta \in \Theta^\star_Z(I)} [\psi_Z(\cdot; \theta)]^{\mathbf{X}}
\end{equation}
in place of the maximizer of~\eqref{eq:per_particle_cost}, with $I = \cL(\mathbf{X})$ and $\Theta^\star_Z(I)$ the maps of Section~\ref{sec:filter}.
For $|Z| \le |I|$ this weight is the likelihood~\eqref{eq:mt_likelihood}.
For $|Z| > |I|$, where~\eqref{eq:mt_likelihood} is zero, it is the value of~\eqref{eq:point_weight} at each map of $\Theta^\star_Z(I)$, summed over those maps.

\begin{proposition}[Dependence of one update on the imputation draws]\label{prop:imputation_inert}
Consider one update of Algorithm~\ref{alg:swarm_lrfs}, and write $\varepsilon = \{\varepsilon^{(i)}_\ell : \ell \notin I^{(i)}\}$ for its imputation draws.
(i) $\theta^{(i)\star}$ and $w^{(i, \theta^{(i)\star})}$ are functions of $\Lambda^{(i)}_{I^{(i)}}$, $Z$, and $R$.
(ii) If $I^{(i)} = I$ for every $i$, the posterior law of the label sets and of $(\Lambda^{(i)}_I)_{i=1}^{N}$ does not depend on $\varepsilon$; with $I = \LL$, $\varepsilon$ is empty and the update is the EnGMF update~\eqref{eq:engmf_update} on $\R^{dL}$ at the maximizer of~\eqref{eq:per_particle_cost}.
(iii) If $I^{(i)} \ne I^{(i')}$ for some $i, i'$, then $\varepsilon$ affects the posterior law of $(\Lambda^{(i)}_{I^{(i)}})_{i=1}^{N}$ only through $B$.
\end{proposition}

Popov and Zanetti~\cite[p.~5, Thm.~2.8]{popov2022} prove that the ensemble Gaussian mixture filter converges to a sequential importance resampling filter as $N \to \infty$ when the observation model is exact and the kernel covariances vanish as $N \to \infty$.
With forecast samples drawn from the dynamics, the limiting sequential importance resampling filter is the bootstrap particle filter~\cite[p.~126]{sarkka2013}.
A scalar normal-reference bandwidth vanishes as $N \to \infty$~\cite[\S2.2]{popov2022}.
Proposition~\ref{prop:large_ensemble} states the vanishing condition for the imputed vectors $\Lambda^{(i)} \in \R^{dL}$, whose sample covariance is~\eqref{eq:blended_moment}.

\begin{proposition}[Large-ensemble limit of the idealized filter]\label{prop:large_ensemble}
Assume:
\begin{enumerate}
\item[(P1)] for some $q > (dL + 4)/2$, the predicted particles are i.i.d.\ with $\Exp\|\Lambda(\mathbf{X})_{\cL(\mathbf{X})}\|^{q} < \infty$, and every imputation draw satisfies $\Exp\|\varepsilon_\ell\|^{q} < \infty$;
\item[(P2)] $r_R I_{d_z} \leq R \leq \bar r_R I_{d_z}$ with $r_R > 0$, and $h$ is differentiable with $\|\nabla h(x)\| \le \bar H$ for every $x \in \X$;
\item[(P3)] $p_D \in (0,1)$, so that the weight~\eqref{eq:point_weight} is positive and finite at every particle;
\item[(P4)] both filters take the same predicted particles after the alignment of line~3, the ancestor map $a$ of~\eqref{eq:resample} is drawn once from the weights~\eqref{eq:idealised_weight} and used by both, and the idealized filter uses Silverman's bandwidth~\eqref{eq:silverman}~\cite{silverman1986}.
\end{enumerate}
Then $\|B\| = \mathcal{O}_{\mathbb{P}}(N^{-2/(dL+4)}) \to 0$, so the covariance hypothesis of~\cite[Thm.~2.8]{popov2022} holds.
Let $\{\mathbf{X}^{(i)}\}_{i=1}^{N}$ be the output of one update of the idealized filter and $\{\mathbf{X}^{(i)}_{\mathrm{PF}}\}_{i=1}^{N}$ the output of one update of the multi-object particle filter of Section~\ref{sec:lrfs_kde} with the weight~\eqref{eq:idealised_weight} in~\eqref{eq:bootstrap_weight} and the ancestor map $a$ of (P4) in place of~\eqref{eq:bootstrap_ancestor}.
Then $\cL(\mathbf{X}^{(i)}) = \cL(\mathbf{X}^{(i)}_{\mathrm{PF}}) = I^{(a(i))}$ for every $i$ and every $N$, and $\|\Lambda(\mathbf{X}^{(i)})_{I^{(a(i))}} - \Lambda(\mathbf{X}^{(i)}_{\mathrm{PF}})_{I^{(a(i))}}\| = \mathcal{O}_{\mathbb{P}}(N^{-1/(dL+4)})$.
The output $\{\mathbf{X}^{(i)}_{\mathrm{PF}}\}_{i=1}^{N}$ involves no imputation draw.
\end{proposition}
When $\cL(\mathbf{X}^{(i)}) = \LL$ for every particle $i$ at every time step, the limiting filter is the fixed-count joint particle filter~\cite[Algorithm~1, eqs.~(8)--(10)]{khan_balch_dellaert_2005}, \cite[Sec.~III]{orton2002bayesian} with the transition kernel $\bfk$ as importance function~\cite[Ch.~7]{sarkka2013} and the likelihood~\eqref{eq:mt_likelihood}.

Proposition~\ref{prop:large_ensemble} does not cover the implemented filter.
The implemented filter evaluates the weight~\eqref{eq:point_weight} at the maximizer of~\eqref{eq:per_particle_cost} and adds $10^{-8} I_{dL}$, a term that does not vanish with $N$, to the kernel covariance~\eqref{eq:silverman} and to $P^{(i, \theta)}$ before the draw in~\eqref{eq:resample}.
Section~\ref{sec:experiments} measures the implemented filter.

\section{Experiments}\label{sec:experiments}

\begin{figure}[!t]
\centering
\includegraphics[width=\columnwidth]{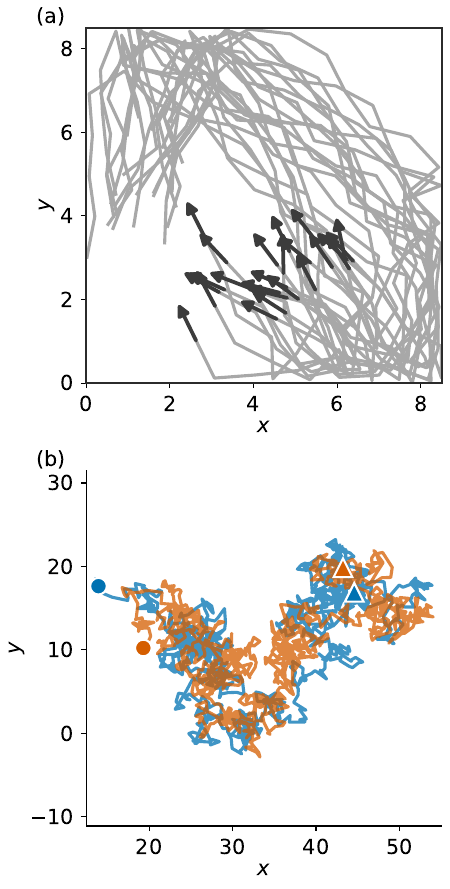}
\caption{Two swarm truth processes. (a) The Vicsek model~\eqref{eq:vicsek}--\eqref{eq:vicsek_position} with twenty-four agents at time step $60$, with arrows along the headings and gray curves over the last twenty steps. (b) The swarm Brownian motion~\eqref{eq:sbm} at $\gamma = -0.05$ with two agents over $1000$ time steps, with a circle at the first step and a triangle at the last.}
\label{fig:truth_processes}
\end{figure}

The experiments measure the tracking error of the SI-LRFS-EnGMF relative to the $\delta$-GLMB-EnGMF as the ensemble size, the coupling between targets, and the detection probability vary.
The Vicsek model of Section~\ref{sec:scenario} runs at six and at twenty-four agents, and the swarm Brownian motion of Section~\ref{sec:sbm} runs at six agents. Table~\ref{tab:protocol} lists the regimes, and Fig.~\ref{fig:truth_processes} draws one realization of each truth process.

\begin{table}[!t]
\caption{Settings of the three regimes. The SI-LRFS-EnGMF and the joint particle filter predict with the transition of the truth process, and the $\delta$-GLMB-EnGMF with the single-agent transition of Sections~\ref{sec:scenario} and~\ref{sec:sbm}.}
\label{tab:protocol}
\centering\small
\setlength{\tabcolsep}{4pt}
\begin{tabular}{@{}lccc@{}}
\hline
 & Vicsek & Vicsek & SBM \\
\hline
truth, initial, birth & Sec.~\ref{sec:scenario} & Sec.~\ref{sec:scenario} & Sec.~\ref{sec:sbm} \\
agents & 6 & 24 & 6 \\
square side $A$ & 4.25 & 8.5 & none \\
label set size $L$ & 10 & 36 & 10 \\
birth probability $r_B$ & 0.025 & 0.1 & 0.025 \\
independent runs & 16 & 8 & 16 \\
$p_S$, $n_B$ & \multicolumn{3}{c}{0.99, 3} \\
$p_D$, $R$ & \multicolumn{3}{c}{0.85, $0.25\, I_2$} \\
time steps $T$, $T_0$ & \multicolumn{3}{c}{110, 10} \\
OSPA order $p$, cutoff $c$ & \multicolumn{3}{c}{2, 3} \\
$\delta$-GLMB-EnGMF & \multicolumn{3}{c}{32 hypotheses, 64 samples per track} \\
\hline
\end{tabular}
\end{table}

In every regime the sensor observes each detected agent's position through the linear-Gaussian likelihood~\cite[Sec.~IV-B]{vo2014labeled}
\begin{equation}\label{eq:vicsek_obs}
g(z|x, \ell) = \cN(z; Hx, R),
\end{equation}
with $H$ the matrix selecting the position coordinates of $x$, the covariance $R$ and detection probability $p_D$ of Table~\ref{tab:protocol}, and no false alarms.
The truth keeps a fixed number of agents.
All three filters run the birth--death law~\eqref{eq:birth_death_kernel} with the rates of Table~\ref{tab:protocol}, whose expected count $n_B r_B / (1 - p_S)$ exceeds the true count in every regime.
All three filters draw the initial label set by including each $\ell \in \LL$ independently with probability $n_B r_B / ((1 - p_S) L)$, and the state of each included $\ell$ from $p_B(\cdot, \ell)$.
The SI-LRFS-EnGMF and the joint particle filter predict with the joint transition of the truth and the $\delta$-GLMB-EnGMF with its restriction to one track, so the SI-LRFS-EnGMF and the $\delta$-GLMB-EnGMF differ in the transition as well as in the density, \eqref{eq:gmm_lrfs} against~\eqref{eq:glmb}.
The joint particle filter has the ensemble size $N$ and initial ensemble of the SI-LRFS-EnGMF, weights particle $i$ by~\eqref{eq:point_weight}, and sets $\mathbf{X}^{(i)} = \mathbf{X}^{(a(i))}$ with $a(i)$ the index of~\eqref{eq:resample}.

A label $\ell$ is declared present when its existence probability $\hat r(\ell)$ of Section~\ref{sec:belief} exceeds $1/2$, and its estimate is the mean of $\Lambda^{(i)}_\ell$ over $\{i : \ell \in I^{(i)}\}$.
The trajectory-mean OSPA of run $s$ is the OSPA distance~\cite{schuhmacher2008} with the order $p$ and cutoff $c$ of Table~\ref{tab:protocol}, averaged over the time steps after the first $T_0$,
\begin{equation}\label{eq:seed_score}
\bar{O}_s = \frac{1}{T - T_0} \sum_{k=T_0+1}^{T} \mathrm{OSPA}_k^{(s)}.
\end{equation}
The $\delta$-GLMB-EnGMF has $64$ samples per track because doubling them to $128$ lowers its across-run mean of $\bar O_s$ by less than $0.02$ in every regime of Table~\ref{tab:protocol}.
The margin on run $s$ is $M_s = \bar O_s^{\delta\text{-GLMB}} - \bar O_s^{\mathrm{SI}}$, positive when the SI-LRFS-EnGMF has the lower $\bar O_s$.
The standard error of the across-run mean of $M_s$ comes from resampling runs with replacement.
All three filters on run $s$ receive the same truth, measurements, and initialization.

\subsection{Vicsek model}\label{sec:scenario}

\begin{figure*}[!t]
\centering
\includegraphics[width=0.9\textwidth]{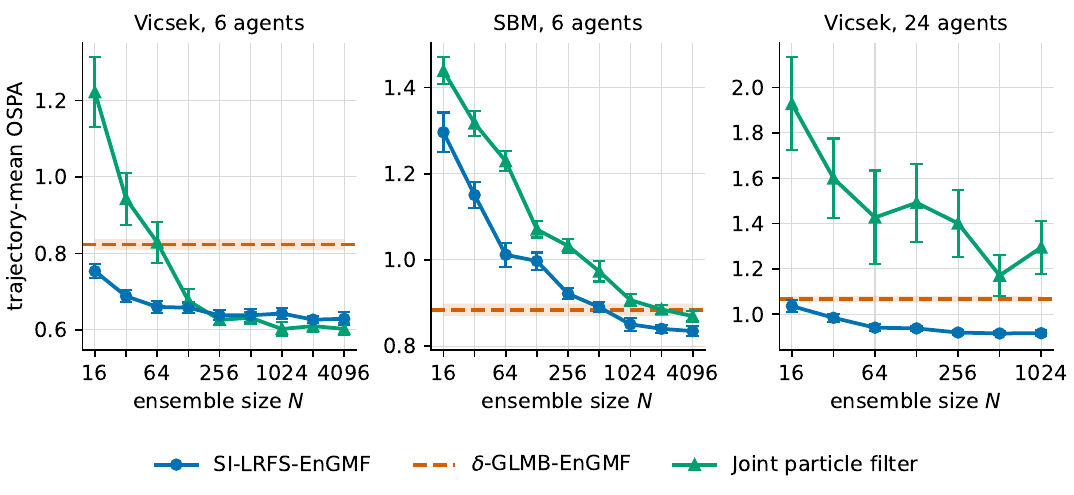}
\caption{Trajectory-mean OSPA~\eqref{eq:seed_score} against the ensemble size $N$ in the three regimes of Table~\ref{tab:protocol}, mean $\pm$ one standard error over the independent runs. The $\delta$-GLMB-EnGMF runs $32$ hypotheses with $64$ samples per track at every $N$.}
\label{fig:n_scaling}
\end{figure*}

The Vicsek model~\cite{vicsek1995} is a simple model of cooperating motion of swarms such as fish and birds.
In it, each agent aligns its heading with the circular mean heading of its neighbors and steps forward at constant speed,
\begin{align}
\varphi_{k+1,i} &= \langle \varphi_k \rangle_{r}(i) + \nu_{k,i}, \label{eq:vicsek}\\
x_{k+1,i} &= x_{k,i} + v_0 \tau u(\varphi_{k+1,i}), \label{eq:vicsek_position}
\end{align}
with $\varphi_{k,i}$ the heading of agent $i$ at time $k$, $\langle \varphi_k \rangle_r(i)$ the circular mean heading over the agents within radius $r$ of agent $i$, $\tau$ the time step, $u(\varphi) = (\cos\varphi, \sin\varphi)^T$, and the heading noise $\nu_{k,i} \sim \mathrm{Unif}[-\nu_0/2, \nu_0/2]$.
The parameters are
\begin{equation}\label{eq:vicsek_params}
v_0 = 2.0, \; r = 3.0, \; \nu_0 = 1.4, \; \tau = 0.5,
\end{equation}
the point of the grid $\nu_0 \in \{0.6, 1.0, 1.4\}$, $v_0 \in \{2.0, 3.0, 4.0\}$ at which the SI-LRFS-EnGMF has the largest OSPA margin over the $\delta$-GLMB-EnGMF.
The agents move in the square $[0, A]^2$ with $A = 4.25$ for six agents and $A = 8.5$ for twenty-four, and the truth and both filters reflect the position at its boundary.
Initial positions and headings are uniform, and newborn states are $\cN((A/2, A/2)^T, (A/4)^2 I_2)$ in position with heading variance $\pi^2/3$.
The SI-LRFS-EnGMF has~\eqref{eq:vicsek}--\eqref{eq:vicsek_position} as its transition kernel on the state $(x, y, \varphi)^T$ per label, with the alignment of Section~\ref{sec:slot_alignment} iterated eight times per time step.
The $\delta$-GLMB-EnGMF predicts each track by~\eqref{eq:vicsek}---\eqref{eq:vicsek_position} with $\langle \varphi_k \rangle_r(i) = \varphi_{k,i}$, since a per-track density has no joint state at which $\langle \varphi_k \rangle_r(i)$ could be evaluated.

The left and right panels of Fig.~\ref{fig:n_scaling} shows the trajectory-mean OSPA of the SI-LRFS-EnGMF below that of the $\delta$-GLMB-EnGMF at every ensemble size $N$, with six agents (left panel) and twenty-four agents (right panel).

The results show that the $\delta$-GLMB-EnGMF fails to capture the cooperating swarm motion for both six and twenty-four agents, that the joint particle filter and the SI-LRFS-EnGMF both converge to the same error for large particle number, and that the SI-LRFS-EnGMF requires significantly fewer particles than the joint particle filter for a large number of agents.

\subsection{Swarm Brownian motion}\label{sec:sbm}

The second experiment varies the coupling $\gamma$ in the drift of a linear diffusion of $n$ agents.
The swarm Brownian motion (SBM) truth is
\begin{equation}\label{eq:sbm}
dx_i = \gamma \sum_{j \neq i} (x_i - x_j)\, dt + \sigma_q\, dW_i,
\end{equation}
with $W_1, \ldots, W_n$ independent standard Brownian motions on $\R^2$, noise scale $\sigma_q$, and signed coupling $\gamma$.
Equation~\eqref{eq:sbm} is the graph Ornstein--Uhlenbeck process of Courgeau and Veraart~\cite{courgeau2022} on the complete graph of the agents, with drift matrix $-\gamma (n - 1)(I - \bar A)$ and $\bar A$ the row-normalized adjacency matrix.
Since $\sum_{j\neq i}(x_i - x_j) = n(x_i - \bar x)$ with $\bar x$ the centroid, the deviation $e_i = x_i - \bar x$ satisfies $de_i = \gamma n\, e_i\, dt + \sigma_q\, d(W_i - \bar W)$ with $\bar W = n^{-1} \sum_i W_i$.
The centroid satisfies $d\bar x = \sigma_q\, d\bar W$, a Brownian motion.
The truth and both filters propagate their transitions by the Euler--Maruyama step~\cite[Alg.~8.1]{sarkka2019} of length $\tau$, whose increment $\sigma_q (W_i(t + \tau) - W_i(t))$ has covariance $\tau \sigma_q^2 I_2$ and under which $e_i$ is multiplied by $1 + \tau\gamma n$ per step.
The state per agent is the position ($d = 2$), so the measurement matrix is the identity $H = I_2$.
The experimental setup is,
\begin{gather}
n = 6, \gamma = -0.05, \quad \sigma_q = 1.0, \quad \tau = 0.5. \label{eq:sbm_settings}
\end{gather}

Initial positions are $\cN((15, 15)^T, 2.5^2 I_2)$ and newborn states $\cN((15, 15)^T, 5^2 I_2)$.
The SI-LRFS-EnGMF's transition kernel is~\eqref{eq:sbm} with the $\gamma$ of the truth, and the $\delta$-GLMB-EnGMF's per-track transition is the random walk
\begin{equation}\label{eq:rw_baseline}
dx = \sigma_q\, dW,
\end{equation}
the diffusion term of~\eqref{eq:sbm}.

\begin{figure}[!t]
\centering
\includegraphics[width=0.9\columnwidth]{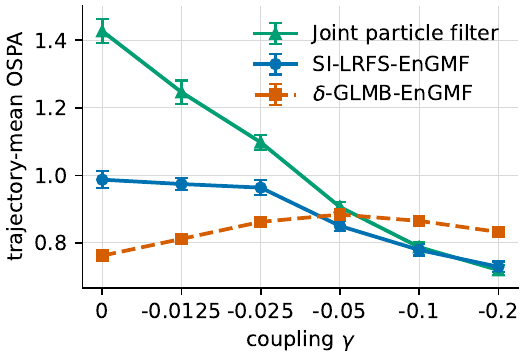}
\caption{Trajectory-mean OSPA~\eqref{eq:seed_score} against the coupling $\gamma$ in the swarm Brownian motion regime of Table~\ref{tab:protocol} with $\gamma$ varied, at $N = 1024$, mean $\pm$ one standard error over the independent runs.}
\label{fig:coupling_ospa}
\end{figure}

Fig.~\ref{fig:coupling_ospa} shows the $\delta$-GLMB-EnGMF with the lower OSPA at $\gamma = 0$, where~\eqref{eq:rw_baseline} is the truth, and the SI-LRFS-EnGMF with the lower OSPA once $|\gamma|$ is large enough, with the margin growing with $|\gamma|$.
The center panel of Fig.~\ref{fig:n_scaling} shows that at the coupling of~\eqref{eq:sbm_settings} the SI-LRFS-EnGMF has the lower OSPA only at the larger ensemble sizes.
This shows that the $\delta$-GLMB-EnGMF is convergent when there is no cooperative motion between targets, but is not exact in the case of a cooperating swarm, and that the SI-LRFS-EnGMF is more stable at representing disjoint states over that of the joint particle filter.

\subsection{Probability of detection}\label{sec:pd_dependence}
\begin{figure}[!t]
\centering
\includegraphics[width=0.9\columnwidth]{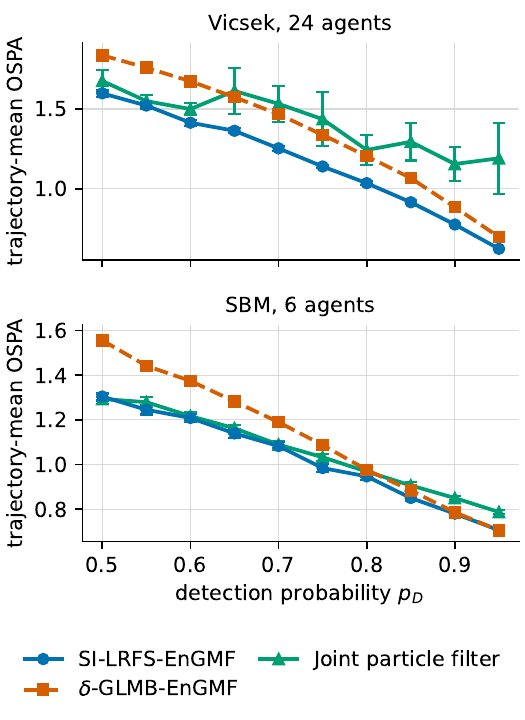}
\caption{Trajectory-mean OSPA~\eqref{eq:seed_score} against the probability of detection $p_D$ in the twenty-four-agent Vicsek regime and the swarm Brownian motion regime of Table~\ref{tab:protocol} with $p_D$ varied, at $N = 1024$, mean $\pm$ one standard error over the independent runs.}
\label{fig:pd_sweep}
\end{figure}

Our third and last experiment reruns the twenty-four-agent Vicsek model and the swarm Brownian motion at $N = 1024$ with the probability of detection $p_D$ varied over the grid of Fig.~\ref{fig:pd_sweep}.
Fig.~\ref{fig:pd_sweep} shows the margin of the SI-LRFS-EnGMF over the $\delta$-GLMB-EnGMF widening as $p_D$ decreases on both regimes.
This means that the SI-LRFS-EnGMF is better able to infer the motion of dependent agents when they move in a cooperating way.

\section{Conclusion}\label{sec:conclusion}\label{sec:discussion}

This work introduced the stochastic-imputation labeled random-finite-set ensemble Gaussian mixture filter (SI-LRFS-EnGMF) for multi-target tracking with cooperative swarms. 
The SI-LRFS-EnGMF constructs a joint kernel-density representation whose cross-target covariance blocks encode inter-agent dependence. 
Consequently, measurements associated with detected targets can update the inferred states of dependent but undetected targets, which is a capability that the $\delta$-GLMB lacks. 

Experiments on Vicsek models and swarm Brownian motion systems demonstrate that preserving joint dynamics and cross-target dependence can substantially improve tracking accuracy, particularly for strongly coupled swarms, larger agent populations, and lower detection probabilities.

The large-ensemble limit of Section~\ref{sec:convergence} holds for the weight~\eqref{eq:idealised_weight}, and whether the implemented filter, with the maximizer of~\eqref{eq:per_particle_cost}, has the same limit is an open question.

Future work would involve more advanced imputation methods, better data alignment, better data association, including clutter models, and a swarm measurement model.
Other future work would involve constructing consistent tracks for individual swarm members.



\section*{Acknowledgments}

A Large Language Model (LLM), Claude Opus 5.5, was used in proof-reading this work. The authors take full responsibility for the contents of the entire work.

\section{Biography Section}

\begin{IEEEbiography}[{\includegraphics[width=1in,height=1.25in,clip,keepaspectratio]{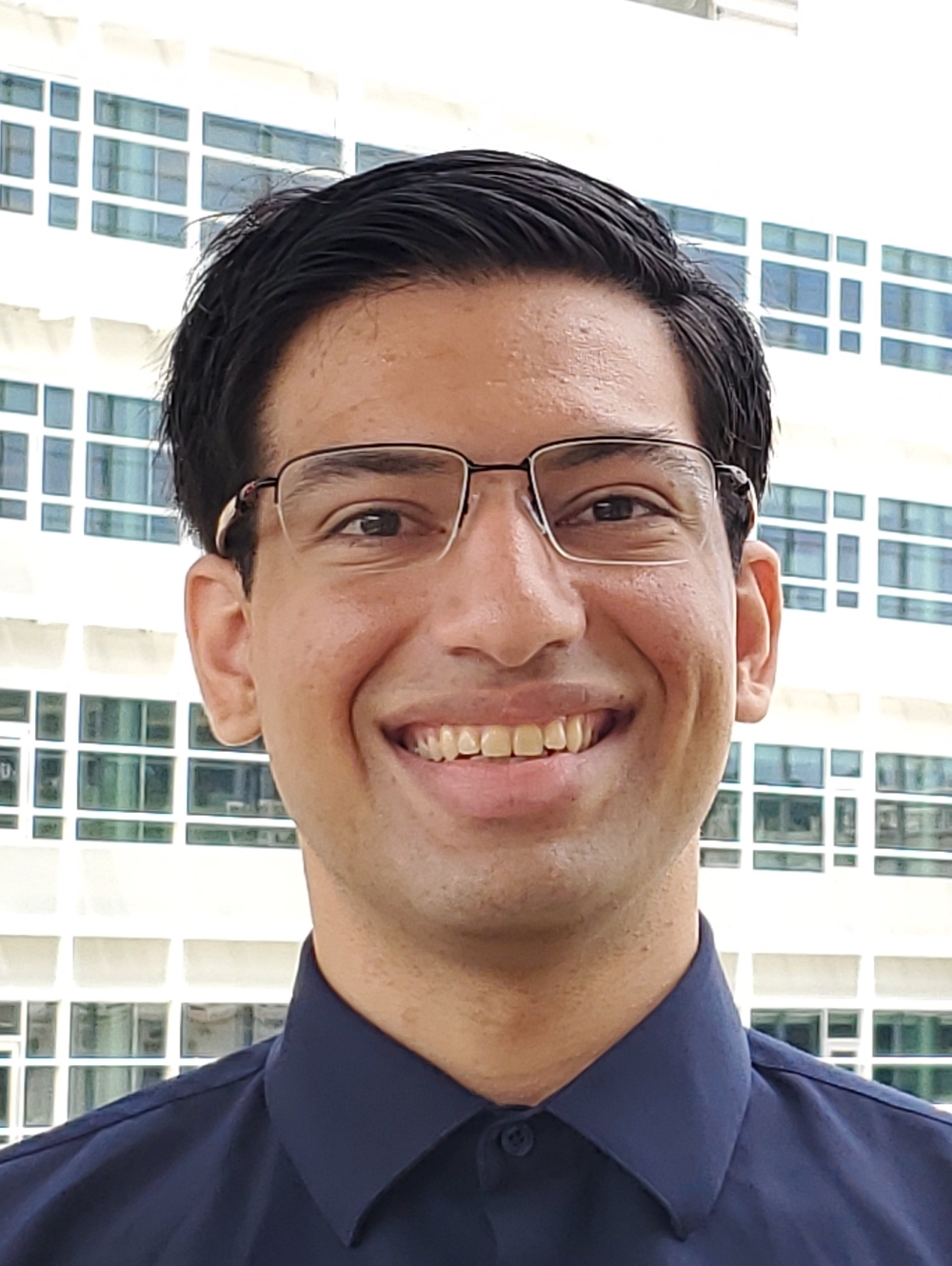}}]{Zain Jabbar}
Zain Jabbar received his B.S. in Mathematics from the University of Hawai'i at M\=anoa, Honolulu, HI.
He is currently a Ph.D. candidate in the Department of Mathematics at the University of Hawai'i at M\=anoa, advised by Dr.\ Andrey A.\ Popov.
His research interests include stochastic filtering, machine learning, and multi-target tracking.
\end{IEEEbiography}

\begin{IEEEbiography}[{\includegraphics[width=1in,height=1.25in,clip,keepaspectratio]{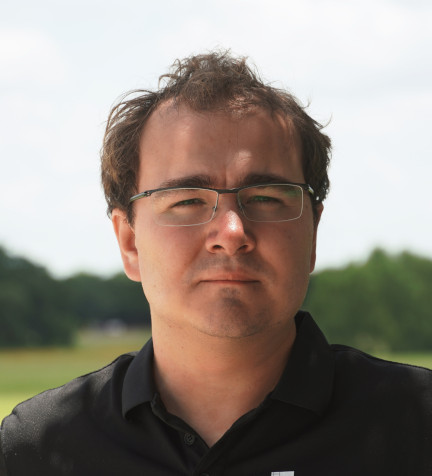}}]{Andrey A. Popov}
Andrey A. Popov received his B.S. in Mathematics from RPI, Troy, NY, and his Ph.D. in Computer Science from Virginia Tech, and was a postdoctoral fellow at the Oden Institute for Computational Engineering and Sciences at the University of Texas at Austin.
He is currently an Assistant Professor at the Department of Information \& Computer Sciences at the University of Hawai'i at Manoa.
He has authored over 40 peer-reviewed papers in journals an conferences including in SIAM SISC and FUSION.
In 2024 and 2026 he was awarded the Jean-Pierre Le Cadre prize at the International Conference on Information Fusion (FUSION).
His primary research focuses on fusing theory-driven and data-driven methods for scientific applications, data assimilation, and aerospace problems. His other research interests include directional statistics, and geographic information systems.
\end{IEEEbiography}

\vfill


\begin{thebibliography}{99}
\bibitem{vo2017}
B.-N.\ Vo, B.-T.\ Vo, and H.\ G.\ Hoang,
``An efficient implementation of the generalized labeled multi-Bernoulli filter,''
\emph{IEEE Trans.\ Signal Process.}, vol.~65, no.~8, pp.~1975--1987, 2017.

\bibitem{vicsek1995}
T.\ Vicsek, A.\ Czir\'{o}k, E.\ Ben-Jacob, I.\ Cohen, and O.\ Shochet,
``Novel type of phase transition in a system of self-driven particles,''
\emph{Phys.\ Rev.\ Lett.}, vol.~75, no.~6, pp.~1226--1229, 1995.

\bibitem{courgeau2022}
V.\ Courgeau and A.~E.~D.\ Veraart,
``Likelihood theory for the graph Ornstein--Uhlenbeck process,''
\emph{Stat.\ Inference Stoch.\ Process.}, vol.~25, pp.~227--260, 2022.

\bibitem{schuhmacher2008}
D.\ Schuhmacher, B.-T.\ Vo, and B.-N.\ Vo,
``A consistent metric for performance evaluation of multi-object filters,''
\emph{IEEE Trans.\ Signal Process.}, vol.~56, no.~8, pp.~3447--3457, 2008.

\bibitem{silverman1986}
B.~W.\ Silverman,
\emph{Density Estimation for Statistics and Data Analysis}.
Chapman \& Hall, 1986.

\bibitem{stephens2000}
M.\ Stephens,
``Dealing with label switching in mixture models,''
\emph{J.\ R.\ Stat.\ Soc.\ B}, vol.~62, no.~4, pp.~795--809, 2000.

\bibitem{celeux1998}
G.\ Celeux,
``Bayesian inference for mixtures: the label-switching problem,''
in \emph{COMPSTAT 98}, R.\ Payne and P.\ Green, Eds.
Heidelberg: Physica, 1998, pp.~227--232.

\bibitem{cover2006}
T.~M.\ Cover and J.~A.\ Thomas,
\emph{Elements of Information Theory}, 2nd~ed.
Wiley, 2006.

\bibitem{dlmf}
F.~W.~J.\ Olver, A.~B.\ Olde Daalhuis, D.~W.\ Lozier, B.~I.\ Schneider, R.~F.\ Boisvert, C.~W.\ Clark, B.~R.\ Miller, B.~V.\ Saunders, H.~S.\ Cohl, and M.~A.\ McClain, eds.,
\emph{NIST Digital Library of Mathematical Functions}, https://dlmf.nist.gov/, accessed 2026-09-18.

\bibitem{nguyen2020}
T.~T.\ Nguyen, H.~D.\ Nguyen, F.\ Chamroukhi, and G.~J.\ McLachlan,
``Approximation by finite mixtures of continuous density functions that vanish at infinity,''
\emph{Cogent Math.\ Stat.}, vol.~7, no.~1, art.~1750861, 2020.

\bibitem{stewart1993}
G.~W.\ Stewart,
``On the early history of the singular value decomposition,''
\emph{SIAM Rev.}, vol.~35, no.~4, pp.~551--566, 1993.

\bibitem{durant2023}
D.\ Durant, A.~A.\ Popov, R.\ Zanetti,
``MCMC EnGMF for sparse data orbit determination,''
in \emph{Proc.\ AAS/AIAA Astrodynamics Specialist Conf.}, 2023, paper AAS~23-356.

\bibitem{cucker_smale_2007}
F.\ Cucker and S.\ Smale,
``Emergent behavior in flocks,''
\emph{IEEE Trans.\ Autom.\ Control}, vol.~52, no.~5, pp.~852--862, 2007.

\bibitem{mahler2007}
R.~P.~S.\ Mahler,
\emph{Statistical Multisource-Multitarget Information Fusion}.
Artech House, 2007.

\bibitem{mahler2014}
R.~P.~S.\ Mahler,
\emph{Advances in Statistical Multisource-Multitarget Information Fusion}.
Artech House, 2014.

\bibitem{vo_singh_doucet_2005}
B.-N.\ Vo, S.\ Singh, A.\ Doucet,
``Sequential Monte Carlo methods for multi-target filtering with random finite sets,''
\emph{IEEE Trans.\ Aerosp.\ Electron.\ Syst.}, vol.~41, no.~4, pp.~1224--1245, 2005.

\bibitem{khan_balch_dellaert_2005}
Z.\ Khan, T.\ Balch, F.\ Dellaert,
``MCMC-based particle filtering for tracking a variable number of interacting targets,''
\emph{IEEE Trans.\ Pattern Anal.\ Mach.\ Intell.}, vol.~27, no.~11, pp.~1805--1819, 2005.

\bibitem{reifler2021}
B.~L.\ Reifler, S.\ Yun, B.~A.\ Jones, R.\ Zanetti,
``Multi-target ensemble Gaussian mixture tracking with sparse observations,''
in \emph{Proc.\ Advanced Maui Optical and Space Surveillance Technologies Conf.\ (AMOS)}, 2021.

\bibitem{sarkka2013}
S.\ S\"arkk\"a,
\emph{Bayesian Filtering and Smoothing},
IMS Textbooks, vol.~3, Cambridge Univ.\ Press, 2013.

\bibitem{gerber2019}
M.\ Gerber, N.\ Chopin, and N.\ Whiteley,
``Negative association, ordering and convergence of resampling methods,''
\emph{Ann.\ Statist.}, vol.~47, no.~4, pp.~2236--2260, 2019.

\bibitem{sarkka2019}
S.\ S\"arkk\"a and A.\ Solin,
\emph{Applied Stochastic Differential Equations}.
Cambridge Univ.\ Press, 2019.

\bibitem{bengtsson2008}
T.\ Bengtsson, P.\ Bickel, B.\ Li,
``Curse-of-dimensionality revisited: collapse of the particle filter in very large scale systems,''
in \emph{Probability and Statistics: Essays in Honor of David A.\ Freedman},
IMS Collections, vol.~2, Institute of Mathematical Statistics, 2008, pp.~316--334.

\bibitem{bickel2008}
P.\ Bickel, B.\ Li, T.\ Bengtsson,
``Sharp failure rates for the bootstrap particle filter in high dimensions,''
in \emph{Pushing the Limits of Contemporary Statistics: Contributions in Honor of Jayanta K.\ Ghosh},
IMS Collections, vol.~3, Institute of Mathematical Statistics, 2008, pp.~318--329.

\bibitem{orton2002bayesian}
M.\ Orton, W.\ Fitzgerald,
``A Bayesian approach to tracking multiple targets using sensor arrays and particle filters,''
\emph{IEEE Trans.\ Signal Process.}, vol.~50, no.~2, pp.~216--223, 2002.

\bibitem{garcia_fernandez_2018}
\'{A}.~F.\ Garc\'{i}a-Fern\'{a}ndez, J.\ L.\ Williams, K.\ Granstr\"{o}m, L.\ Svensson,
``Poisson multi-Bernoulli mixture filter: direct derivation and implementation,''
\emph{IEEE Trans.\ Aerosp.\ Electron.\ Syst.}, vol.~54, no.~4, pp.~1883--1901, 2018.

\bibitem{yang2024}
Y.\ Yang, L.\ Pu, X.\ Li, X.\ Liu,
``Cluster target tracking based on multi-sensor continuous-discrete PMBM filter,''
\emph{Expert Syst.\ Appl.}, 2024.

\bibitem{wand_jones_1995}
M.~P.\ Wand, M.~C.\ Jones,
\emph{Kernel Smoothing}.
Chapman \& Hall, 1995.

\bibitem{vo_vo_2013}
B.-T.\ Vo, B.-N.\ Vo,
``Labeled random finite sets and multi-object conjugate priors,''
\emph{IEEE Trans.\ Signal Process.}, vol.~61, no.~13, pp.~3460--3475, 2013.

\bibitem{kuhn1955}
H.~W.\ Kuhn,
``The Hungarian method for the assignment problem,''
\emph{Nav.\ Res.\ Logist.\ Q.}, vol.~2, no.~1--2, pp.~83--97, 1955.

\bibitem{vo2024overview}
B.-N.\ Vo, B.-T.\ Vo, T.~T.~D.\ Nguyen, and C.\ Shim,
``An overview of multi-object estimation via labeled random finite set,''
\emph{IEEE Trans.\ Signal Process.}, vol.~72, 2024, doi:10.1109/TSP.2024.3472068.

\bibitem{challa2011}
S.\ Challa, M.~R.\ Morelande, D.\ Mu\v{s}icki, and R.~J.\ Evans,
\emph{Fundamentals of Object Tracking}.
Cambridge University Press, 2011.

\bibitem{barshalom2001}
Y.\ Bar-Shalom, X.~R.\ Li, and T.\ Kirubarajan,
\emph{Estimation with Applications to Tracking and Navigation}.
Wiley, 2001.

\bibitem{thomas2018}
D.~J.\ Thomas, K.~M.\ Nastasi, K.\ Schroeder, and J.~T.\ Black,
``Autonomous multi-phenomenology space domain sensor tasking and adaptive estimation,''
in \emph{Proc.\ 21st Int.\ Conf.\ Information Fusion (FUSION)}, 2018.

\bibitem{vo2014labeled}
B.-N.\ Vo, B.-T.\ Vo, and D.\ Phung,
``Labeled random finite sets and the Bayes multi-target tracking filter,''
\emph{IEEE Trans.\ Signal Process.}, vol.~62, no.~24, pp.~6554--6567, 2014.

\bibitem{yun2022}
S.\ Yun, R.\ Zanetti, and B.~A.\ Jones,
``Kernel-based ensemble Gaussian mixture filtering for orbit determination with sparse data,''
\emph{Adv.\ Space Res.}, 2022.

\bibitem{popov2022}
A.~A.\ Popov and R.\ Zanetti,
``An adaptive covariance parameterization technique for the ensemble Gaussian mixture filter,''
arXiv:2212.10323; version submitted to \emph{SIAM J.\ Sci.\ Comput.}, January 13, 2024.
The result cited here is the theorem titled ``EnGMF convergence'' (Thm.~2.8 in that version; renumbered across arXiv revisions).

\bibitem{popov2024}
A.~A.\ Popov and R.\ Zanetti,
``The ensemble Epanechnikov mixture filter,''
arXiv:2408.11164, 2024.

\bibitem{anderson1999monte}
J.~L.\ Anderson and S.~L.\ Anderson. 
``A Monte Carlo implementation of the nonlinear filtering problem to produce ensemble assimilations and forecasts.'' \emph{Monthly weather review} 127.12 (1999): 2741-2758.


\end{thebibliography}
\end{document}